 \documentclass[final,5p,times,twocolumn,authoryear]{elsarticle}

\usepackage{amssymb}
\usepackage{amsmath}
\usepackage{lipsum}
\usepackage{multirow}
\usepackage{url}
\usepackage{aas_macros}

\journal{Astronomy $\&$ Computing}

\begin{document}

\begin{frontmatter}



\title{Mercury-Ar$\chi$es: a high-performance n-body code for planet formation studies}


\author[1,2]{Diego Turrini}
\affiliation[1]{organization={INAF - Osservatorio Astrofisico di Torino},
            addressline={Strada Osservatorio 20}, 
            city={Pino Torinese},
            postcode={10020}, 
            state={},
            country={Italy}}
\affiliation[2]{organization={ICSC – National Research Centre for High Performance Computing, Big Data                 and Quantum Computing},
                addressline={Via Magnanelli 2},
                city={Casalecchio di Reno},
                postcode={40033},
                state={},
                country={Italy}}

\author[3]{Sergio Fonte}
\affiliation[3]{organization={INAF - Istituto di Astrofisica e Planetologia Spaziali},
            addressline={Via Fosso del Cavaliere 100}, 
            city={Rome},
            postcode={00133}, 
            state={},
            country={Italy}}

\author[2,3]{Romolo Politi}

\author[1,2]{Danae Polychroni}

\author[3]{Scig\'e J. Liu}

\author[2,4]{Paolo Matteo Simonetti}
\affiliation[4]{organization={INAF - Osservatorio Astronomico di Trieste},
            addressline={Via G.B. Tiepolo 11}, 
            city={Trieste},
            postcode={34143}, 
            state={},
            country={Italy}}

\author[5]{Simona Pirani}
\affiliation[5]{organization={Sharkmob},
            city={Malmö},
            country={Sweden}}

\begin{abstract}

Forming planetary systems are populated by large numbers of gravitationally interacting planetary bodies, spanning from massive giant planets to small planetesimals akin to present-day asteroids and comets. All these planetary bodies are embedded in the gaseous embrace of their native protoplanetary disks, and their interactions with the disk gas play a central role in shaping their dynamical evolution and the outcomes of planet formation. These factors make realistic planet formation simulations extremely computationally demanding, which in turn means that accurately modeling the formation of planetary systems requires the use of high-performance methods. The planet formation code {\sc Mercury-Ar$\chi$es} was developed to address these challenges and, since its first implementation, has been used in multiple exoplanetary and Solar System studies. {\sc Mercury-Ar$\chi$es} is a parallel n-body code that builds on the widely used {\sc Mercury} code and is capable of modeling the growth and migration of forming planets, the interactions between planetary bodies and the disk gas, as well as the evolving impact flux of planetesimals on forming planets across the different stages of their formation process. In this work we provide the up-to-date overview of its physical modeling capabilities and the first detailed description of its high-performance implementation based on the OpenMP directive-based parallelism for shared memory environments, to harness the multi-thread and vectorization features of modern processor architectures.
\end{abstract}




\end{frontmatter}




\section{Introduction}
\label{introduction}

The {\sc Mercury} n-body simulation code \citep{chambers1999} has been the tool of choice for planetary dynamics and terrestrial planet formation studies for decades, enabling more than 1600 investigations to date, and is still widely used in the planetary and exoplanetary science communities. Its open nature, intuitive structure, and clear programming style favored the branching of new versions, such as {\sc SMERCURY} \citep{lissauer2012} for the study of spin-orbit interactions or {\sc Mercury-T} \citep{bolmont2015} for the study of tides, and the development of libraries implementing new modeling capabilities, like the symplectic mapping of binary systems (DPI, \citealt{DPI2015,Nigioni2025} based on \citealt{chambers2002}) and the Yarkovsky-O'Keefe-Radzievskii-Paddack effect \citep[][]{fenucci2022}. {\sc Mercury} also serves as the n-body engine for the NGPPS planetary population synthesis model \citep{Emsenhuber2021}.

New n-body simulation packages have been developed by the community in recent years, introducing parallel or GPU-computing capabilities to allow for addressing more complex dynamical and collisional systems: \texttt{REBOUND} \citep[][]{rein2012}, \texttt{GENGA} \citep[][]{grimm2014} and \texttt{GENGA II} \citep[][]{grimm2022}, and \texttt{GLISSE} \citep[][]{zhang2022}. No n-body current code, however, is optimized for the study of the earliest phases in the life of planetary systems, when forming planets and planetesimals are embedded into their native protoplanetary disk and evolve together with it. In this work we present the first complete and self-contained description of the modeling and computational features Mercury-Ar$\chi$es, an n-body planet formation code combining an high-performance implementation of the hybrid symplectic algorithm of the {\sc Mercury} code with a library allowing to model the main physical processes shaping the formation of planetary systems within their native protoplanetary disks.

The modeling features provided by {\sc Mercury-Ar$\chi$es} allow for high-resolution explorations of the interactions between forming planets and the protoplanetary disks that surround them, and for the detailed characterization of how the planets alter the environment in which they are forming and how these alterations affect the planets themselves. The growth of massive planets is known to alter the dynamical equilibrium of the planetesimal disks embedded in the native circumstellar disks \citep{safronov1972,weidenschilling1998,turrini2011}, causing the planetesimals to become dynamically excited and start intense collisional cascades \citep{weidenschilling2001,turrini2012,turrini2019} long before the protoplanetary disks transition into debris disks. The dynamical excitation of the planetesimal disk results in its compositional remix and the transport of volatile elements across different disk regions \citep{safronov1972,weidenschilling2001,turrini2014b,turrini2015}. This excitation process does not depend solely on planetary perturbations but is enhanced by disk gravity and countered by aerodynamic drag \citep{weidenschilling1977,ward1981,nagasawa2000,nagasawa2019}. Notwithstanding the critical role played by these processes in shaping the earliest phases in the life of planetary systems, they are still under-explored due to the computational challenges their modeling presents \citep[e.g.][and references therein]{turrini2015,turrini2018,turrini2022,Feinstein2025}.

{\sc Mercury-Ar$\chi$es} has allowed for the first self-consistent studies of these processes in the circumstellar disks surrounding young stars \citep{turrini2019,bernabo2022,polychroni2025} and the Sun \citep{sirono2025}. The capability of resolving the concurrent effects of planetary perturbations, aerodynamic drag and disk gravity, while tracking the source regions of the planetesimals, recently allowed for quantifying the dynamical transport and implantation of primordial comets into the inner Solar System and the duration of their collisional cascade responsible for the formation of chondrules \citep{sirono2025}. In parallel, the possibility of characterizing the role of disk gravity as a function of the physical property of the disk allowed to explore how the intensity of the transport and remixing is affected by the observational uncertainty on the masses of protoplanetary disks and of their embedded planets \citep{polychroni2025}).

The dynamical excitation of the planetesimals affects the very planets that stir the planetesimal disk \citep{turrini2015,shibata2019,shibata2020}, enhancing the flux of impactors that hit the planets during their growth and migration. The detailed characterization of the dynamical state of the protoplanetary disk and of the temporal evolution of the physical and gravitational cross-sections of the growing planets of {\sc Mercury-Ar$\chi$es} allowed for the realistic quantification of the accretion of planetesimals by forming planets during the growth and migration across the native disks \citep{turrini2021,turrini2023,polychroni2025}. The record of their source regions allowed in turn to quantify the accretion efficiency across the different compositional regions of protoplanetary disk and the resulting composition of the young planets \citep{turrini2021,pacetti2022,fonte2023,polychroni2025}.

While the fundamental aspects of the modeling features of {\sc Mercury-Ar$\chi$es} were introduced in \citet{turrini2019} and \citet{turrini2021}, no self-contained and updated description of its modeling algorithms is available in the literature. As an example, \citet{turrini2019} adopted the gas drag treatment from \citet{brasser2007}, which was later found to implicitly hard-wire a number of fixed disk parameters into its numerical constants and to present possible discontinuities between the different drag regimes \citep{stoyanovskaya2020}, resulting in its substitution with a more general implementation (see Sect. \ref{sect-protoplanetary_disk}) after the study of \citet{turrini2021}. Similarly, while \citet{turrini2019} and \citet{turrini2021} reported that {\sc Mercury-Ar$\chi$es} was based on a parallel version of the hybrid symplectic algorithm of the {\sc Mercury} code, no details were provided on the adopted optimization and parallelization approach since they were beyond the scope of those studies.

In the following we fill these gaps by first describing {\sc Mercury+}, the parallel n-body engine of {\sc Mercury-Ar$\chi$es} (Sect. \ref{sect-nbody_engine}) and the updated multi-physics engine for planet formation studies (Sect. \ref{sect-physical_engine}). Finally, we illustrate the performance and scalability of {\sc Mercury-Ar$\chi$es} on realistic, non-idealized use cases based on previous investigations using both consumer-grade and cluster infrastructures {Sect. \ref{sect:performance}).}

\section{{\sc Mercury+}, the n--body engine of {\sc Mercury-Ar$\chi$es}}\label{sect-nbody_engine}

{\sc Mercury+} builds on {\sc Mercury} 6 \citep{chambers1999} and provides an high-performance implementation of its hybrid symplectic integrator. This design choice is motivated by the desirable properties of this integration schemes for planet formation studies, since it combines the computational efficiency of symplectic algorithms with the capability of resolving close encounters and collisions among planetary bodies. 

Symplectic algorithms \citep{wisdom1991,kinoshita1991} are a family of leapfrog integration schemes that are about an order of magnitude faster than conventional n-body algorithms and allow to simulate the long-term evolution of planetary systems without undergoing secular accumulation of the energy error. To improve the accuracy and numerical stability of the Keplerian evolution of the planetary orbits, {\sc Mercury-Ar$\chi$es} interfaces with the {\sc WHFAST} library of the {\sc REBOUND} code \citep{rein2015} in place of {\sc Mercury}'s original subroutines for the computation of the Keplerian drift. The seamless interface between {\sc Mercury} (FORTRAN77) and {\sc WHFAST} (C99) is guaranteed by a Fortran95 wrapper.

The energy conservation property of symplectic algorithms is guaranteed as long as the hierarchical structure of the dynamical system is preserved. For planetary systems this means that the dynamical evolution of the bodies is dominated by the gravitational field of the host star, a condition that is violated during close encounters between planetary bodies. During close encounters, Mercury's hybrid symplectic algorithm switches to the high-precision non-symplectic Bulirsh--Stoer integration scheme to accurately reproduce the dynamical evolution of the involved bodies while their gravitational interactions are comparable in magnitude to the gravitational pull of the star \citep{chambers1999}. 

\subsection{High-performance features of the n-body engine}

The hybrid symplectic scheme of {\sc Mercury+} has been parallelized and vectorized with OpenMP to take advantage of the SIMD (single instruction, multiple data) capabilities of modern multi-core architectures. The HPC implementation of {\sc Mercury+} is based on the following design philosophy:
\begin{itemize}
    \item we preserved the original structure of {\sc Mercury} and adopted solutions that maximize the interoperability with legacy codes and libraries designed to work with it, to facilitate their integration with {\sc Mercury+}. This choice led to the adoption of the directive-based OpenMP model for the porting to HPC;
    \item we focused the development on value-safe parallelization solutions to enforce reproducibility between the parallel and serial versions of the code. This design choice has been adopted to facilitate the debugging and validation of existing and future parallel sections of the code;
    \item we explicitly instructed the compiler to combine parallelization and vectorization whenever possible to provide consistent performances across different compilers. Analogously, we implemented documented bug fixes to minimize the dependencies on runtime environments and provide the users with a robust and flexible n--body integrator;
\end{itemize}

We provide details on the implementation of this design philosophy in the following sections. The list of parallel subroutines within {\sc Mercury+} is the following: {\sc mce\_box, mce\_hill, mce\_snif, mco\_b2h, mco\_dh2h, mco\_h2b, mco\_h2dh, mco\_iden, mdt\_bs1, mdt\_bs2, mdt\_hy, mfo\_drct, mfo\_usr, mfo\_hy, mfo\_hkce}. The selection of the subroutines to parallelize was based on the cumulative profiling outputs obtained through {\sc Gprof} across all science cases where we took advantage of {\sc Mercury-Ar$\chi$es} \citep{turrini2019,turrini2021,turrini2023,bernabo2022,polychroni2025,sirono2025}.

From a technical point of view the subroutine {\sc mdt\_bs1}, implementing the general case of the Bulirsh-Stoer integrator in {\sc Mercury}, is not used directly by {\sc Mercury-Ar$\chi$es} as the hybrid symplectic integrator makes use of {\sc mdt\_bs2}, an implementation of the Bulirsh-Stoer integrator specific for the use of conservative forces. However, since {\sc mdt\_bs1} and {\sc mdt\_bs2} share a common structure, we parallelized both for future use. In the following, we provide illustrative examples of the optimizations introduced in {\sc Mercury+}, while Sect. \ref{sect:performance} will showcase typical use cases and workloads of {\sc Mercury-Ar$\chi$es}.

\subsubsection{Header, threading and scheduling}

All subroutines that take advantage of OpenMP parallelization start with the following header:

\begin{verbatim}
!$    use OMP_LIB
!$    include 'openmp.inc'
\end{verbatim} 

Both statements are inserted as optional programming commands, meaning that they are treated as comments when the source code is compiled serially. The explicit inclusion of the {\sc use omp\_lib} statement ensures portability and consistent behavior across different compilers/OpenMP implementations.

The inclusion of the file {\sc openmp.inc} imports into the subroutines the declarations of three variables controlling the parallel execution of the code at runtime:
\begin{itemize}
    \item {\sc n\_th}, an integer constant setting the number of threads to be used when running in parallel;
    \item {\sc omp\_dynflag}, a boolean constant that allows for switching between static scheduling and dynamic scheduling in managing the parallel execution;
    \item {\sc npar\_th}, an integer constant specifying the minimum number of body for parallel execution in case of conditional parallelization (see below).
\end{itemize}

At the beginning of each subroutine containing OpenMP parallel directives, the {\sc n\_th} and {\sc omp\_dynflag} parameters are used to configure the runtime environment in terms of threading and scheduling by calling the relevant subroutines provided by the OpenMP library as:

\begin{verbatim}
!$    call OMP_SET_NUM_THREADS(n_th)
!$    call OMP_SET_DYNAMIC(omp_dynflag)
\end{verbatim}

With GNU compilers we experienced loss of performances when selecting the dynamic scheduler, while with Intel compilers the two choices did not show significant differences in terms of performances. Performance-wise, the safest choice for {\sc omp\_dynflag} is therefore the static scheduler. The parameter {\sc npar\_th} is used instead whenever parallel instructions are used on a subset of bodies to prevent parallel functionalities from being used in situations where insufficient computational load may cause them to hinder performance, e.g. during the resolution of close encounters:

\begin{verbatim}
!$omp  parallel do simd default(shared) 
!$omp& private(tmp1,tmp2) if(nbod>npar_th)
      do k = 2,nbod
        tmp1 = x0(1,k)*x0(1,k)+x0(2,k)*x0(2,k)
     &  +x0(3,k)*x0(3,k)
        tmp2 = v0(1,k)*v0(1,k)+v0(2,k)*v0(2,k)
     &  +v0(3,k)*v0(3,k) 
        xscal(k) = 1.d0/tmp1
        vscal(k) = 1.d0/tmp2
      end do
!$omp end parallel do simd 
\end{verbatim}

\subsubsection{Parallel, Parallel SIDM and Critical regions}

All loops and array operations that can benefit from parallelism have been enclosed in {\sc parallel do} and {\sc parallel workshare} directives, as in the following code examples:

\begin{verbatim}
!$omp parallel do default(shared) private(temp)
      do j = 2, nbod
        xmin(j) = min (x0(1,j), x1(1,j))
        xmax(j) = max (x0(1,j), x1(1,j))
        ymin(j) = min (x0(2,j), x1(2,j))
        ymax(j) = max (x0(2,j), x1(2,j))
c
c If velocity changes sign, do an interpolation
        if ((v0(1,j).lt.0.and.v1(1,j).gt.0).or.
     &   (v0(1,j).gt.0.and.v1(1,j).lt.0)) then
         temp=(v0(1,j)*x1(1,j)-v1(1,j)*x0(1,j)
     &   -0.5d0*h*v0(1,j)*v1(1,j))/(v0(1,j)
     &   -v1(1,j))
         xmin(j) = min (xmin(j),temp)
         xmax(j) = max (xmax(j),temp)
        end if
c
        if ((v0(2,j).lt.0.and.v1(2,j).gt.0).or.
     &   (v0(2,j).gt.0.and.v1(2,j).lt.0)) then
         temp=(v0(2,j)*x1(2,j)-v1(2,j)*x0(2,j)
     &   -0.5d0*h*v0(2,j)*v1(2,j))/(v0(2,j)
     &   -v1(2,j))
         ymin(j) = min (ymin(j),temp)
         ymax(j) = max (ymax(j),temp)
        end if
      end do
!$omp end parallel do
\end{verbatim}

and 

\begin{verbatim}
!$omp parallel workshare
      ejcount=count(stat(:).lt.0)
!$omp end parallel workshare
      if (ejcount.gt.0) ejflag = 1    
\end{verbatim}

To maximize the performance consistency across different platforms and compilers, all parallel loops that can benefit from both parallelization and vectorization have been enclosed within {\sc parallel do simd} directives:

\begin{verbatim}
!$omp parallel do simd default(shared)
      do j = 2, nbod
        v(1,j)=v(1,j)+hby2*(angf(1,j)
     &   +ausr(1,j)+a(1,j))
        v(2,j)=v(2,j)+hby2*(angf(2,j)
     &   +ausr(2,j)+a(2,j))
        v(3,j)=v(3,j)+hby2*(angf(3,j)
     &   +ausr(3,j)+a(3,j))
      end do
!$omp end parallel do simd    
\end{verbatim}

The rationale of this choice is illustrated by the different philosophies of the two mainstream Intel and GNU suite of compilers. While Intel compilers implicitly analyze and vectorize parallel loops to maximize performance, GNU compilers disable implicit vectorization within parallel loops unless explicitly instructed to do so to minimize race conditions and the risk of producing incorrect results. The inclusion of the {\sc parallel do simd} directives instructs the compiler that it is safe to combine parallelization and SIMD instructions.

During the profiling and assessment of the parallelization potential of the {\sc Mercury} n-body library, we identified two code regions that can prove computationally intensive in planet formation simulations but where the structure of the algorithms implemented in {\sc Mercury} prevents their straightforward parallelization. The first code region is associated with the identification of pairs of bodies undergoing close encounters during the timestep and for which the use of the Bulirsh-Stoer integrator is required:
\begin{verbatim}
c If minimum separation is small enough,
c flag this as a possible encounter
            temp = min (d0,d1,d2min)
            if (temp.le.rc2) then
!$omp critical(ce_snif)
              ce(i) = 2
              ce(j) = 2
              nce = nce + 1
              ice(nce) = i
              jce(nce) = j
!$omp end critical(ce_snif)
            end if    
\end{verbatim}

The second code region is associated with the computation of the forces acting between such pairs of bodies during the close encounters with the Bulirsh-Stoer integrator, and could incur in race conditions in those cases when a single body has multiple close encounters during a timestep:

\begin{verbatim}
!$omp critical(fo_hkce)
          faci = tmp2 * m(i)
          facj = tmp2 * m(j)
          a(1,j) = a(1,j)  -  faci * dx
          a(2,j) = a(2,j)  -  faci * dy
          a(3,j) = a(3,j)  -  faci * dz
          a(1,i) = a(1,i)  +  facj * dx
          a(2,i) = a(2,i)  +  facj * dy
          a(3,i) = a(3,i)  +  facj * dz
!$omp end critical(fo_hkce)    
\end{verbatim}

In line with our philosophy of preserving the original structure of Mercury and the interoperability with legacy codes and libraries, we analyzed these code regions and verified that the issue could be efficiently addressed with the introduction of {\sc critical} directives to take advantage of the sparse nature of close encounters and, consequently, access to the critical regions. While the two {\sc critical} regions are located in different subroutines and there is currently no risk of race conditions, to prevent the chance of this occurrence in future updates we implemented them as named {\sc critical} regions (see above).

\subsubsection{Additional upgrades and features debugging}

The source code of {\sc Mercury+} also contains the following differences with respect to {\sc Mercury} \citep{chambers1999}. First, we implemented the bug fix reported in \citet{desouza2008} by explicitly initializing the {\sc stat} variable in the subroutine {\sc mio\_in}. The {\sc stat} variable keeps track of the status of the bodies during the simulations and its explicit initialization prevents the  compiler-dependent possibility of bodies being removed in a non-physical manner during or at the restart of simulations.

Second, we modified the loops computing the total momentum of the system in {\sc mco\_dh2h} and {\sc mdt\_hy} and the loop computing the gravitational potential energy in {\sc mxx\_en} to run only over the number of massive bodies rather than over the total number of bodies. This simple solution proved to be more effective than parallelizing the loops for the use cases investigated with the code \citep{turrini2019,turrini2021,bernabo2022,turrini2023,polychroni2025,sirono2025}. According to tests, parallelizing these loops is advised for simulations including more than $10^4$ massive particles.

Third, we upgraded the {\sc mxx\_sort} subroutine to support arrays up to $\approx8\times10^5$ elements instead of the original $\approx3\times10^4$ to prevent spurious behaviors in large simulations; the upgrade extends the gap sequence of Shellsort in the {\sc incarr} variable up to the 12th term following Knuth's method ($x=(3^n-1)/2$, \citealt{knuth1973}):
\begin{verbatim}
      data incarr/1,4,13,40,121,364,1093,3280,
     & 9841,29524,88573,265720/
\end{verbatim}

Finally, we modified the computation of Newtonian forces in the subroutine {\sc mfo\_drct} to branch automatically depending on whether the code is running in parallel or not, taking advantage of conditional directives. When the code is run serially, the variable {\sc openmpflag} is set to {\sc .FALSE.} and the original and more efficient triangular loop by {\sc Mercury} is used. When the code is run in parallel, the OpenMP directive setting {\sc openmpflag} to {\sc .TRUE.} is activated and a rectangular loop is used instead: 

\begin{verbatim}
      data openmpflag/ .false. /
!$    openmpflag=.true.
      if (i0.le.0) i0 = 2
      if (openmpflag.eqv..false.) then
        do i = i0, nbig
          do j = i + 1, nbod
            dx = x(1,j) - x(1,i)
            dy = x(2,j) - x(2,i)
            dz = x(3,j) - x(3,i)
            s2 = dx*dx + dy*dy + dz*dz
            rc = max(rcrit(i),rcrit(j))
            rc2 = rc * rc
            if (s2.ge.rc2) then
              s_1 = 1.d0 / sqrt(s2)
              tmp2 = s_1 * s_1 * s_1
            else if (s2.le.0.01*rc2) then
              tmp2 = 0.d0
            else
              s_1 = 1.d0 / sqrt(s2)
              s   = 1.d0 / s_1
              s_3 = s_1 * s_1 * s_1
              q = (s - 0.1d0*rc) / (0.9d0 * rc)
              q2 = q  * q
              q3 = q  * q2
              q4 = q2 * q2
              q5 = q2 * q3
              tmp2=(10.d0*q3-15.d0*q4+6.d0*q5)
     &        *s_3
            end if
            faci = tmp2 * m(i)
            facj = tmp2 * m(j)
            a(1,j) = a(1,j)  -  faci * dx
            a(2,j) = a(2,j)  -  faci * dy
            a(3,j) = a(3,j)  -  faci * dz
            a(1,i) = a(1,i)  +  facj * dx
            a(2,i) = a(2,i)  +  facj * dy
            a(3,i) = a(3,i)  +  facj * dz
          end do
        end do
      else
!$omp  parallel do shared(x,rcrit,i0,nbig,nbod,m,a)
!$omp& private(i,j,dx,dy,dz,s2,rc,rc2,s_1,
!$omp& tmp2,s,s_3,q,q2,q3,q4,q5,faci,facj)
        do j = i0, nbod
          do i = i0, nbig
            if (i.ne.j) then
              dx = x(1,j) - x(1,i)
              dy = x(2,j) - x(2,i)
              dz = x(3,j) - x(3,i)
              s2 = dx*dx + dy*dy + dz*dz
              rc = max(rcrit(i), rcrit(j))
              rc2 = rc * rc
c
              if (s2.ge.rc2) then
                s_1 = 1.d0 / sqrt(s2)
                tmp2 = s_1 * s_1 * s_1
              else if (s2.le.0.01*rc2) then
                tmp2 = 0.d0
              else
                s_1 = 1.d0 / sqrt(s2)
                s   = 1.d0 / s_1
                s_3 = s_1 * s_1 * s_1
                q = (s-0.1d0*rc)/(0.9d0*rc)
                q2 = q  * q
                q3 = q  * q2
                q4 = q2 * q2
                q5 = q2 * q3
                tmp2=(10.d0*q3-15.d0*q4+6.d0*q5)
     &          *s_3
              end if
              faci = tmp2 * m(i)
              a(1,j) = a(1,j)  -  faci * dx
              a(2,j) = a(2,j)  -  faci * dy
              a(3,j) = a(3,j)  -  faci * dz
	          end if
          end do
        end do
!$omp end parallel do
      end if    
\end{verbatim}

While theoretically less efficient than other solutions, e.g. collapsing the nested loops into a single one over pairs of bodies, this approach consistently provided the best performance.

\section{{Multi-physics engine of \sc Mercury-Ar$\chi$es}}\label{sect-physical_engine}

The second component of {\sc Mercury-Ar$\chi$es} is the {\sc Ar$\chi$es} library, which allows for modeling the physical processes involved in planet formation and the interactions of the planetary bodies with the surrounding protoplanetary disk. The library is composed of the following thematic Fortran95 modules and the main FORTRAN77 library:
\begin{itemize}
    \item {\sc arxes\_common.f90}, a module library of astrophysical and conversion constants and mathematical functions used by the other modules;
    \item {\sc arxes\_disk.f90}, the module library containing the variables and initialization functions used for the computation of the interactions between the disk gas and the planetary bodies (Sect. \ref{sect-protoplanetary_disk});
    \item {\sc arxes\_growth.f90}, the module library containing the variables and subroutines implementing the mass growth and planetary radius evolution of forming planets (Sect. \ref{sec:planetary_growth});
    \item {\sc arxes\_migration.f90}, the module library containing the variables and subroutines implementing the orbital migration of forming planets (Sect. \ref{sec:orbital_migration}).
    \item {\sc arxes.for}, the main library, containing also modified versions of {\sc Mercury} subroutines in line with the preservation philosophy of {\sc Mercury+} (Sects. \ref{sect-protoplanetary_disk},\ref{sec:planetary_growth},\ref{sec:orbital_migration}).
\end{itemize}
In the following we detail the modeling algorithms and computational solutions adopted in the {\sc Ar$\chi$es} library.


\subsection{Protoplanetary disks and dynamical effects of gas}\label{sect-protoplanetary_disk}

The protoplanetary disk modeling assumes a viscous disk in steady state \citep{lindenbell1974,andrews2010} parametrized as 
\begin{equation}
\Sigma_{gas}(r) = \Sigma_{0}\left(\frac{r}{R_{C}}\right)^{-\gamma} \exp{\left[-\left(\frac{r}{R_{C}}\right)^{2-\gamma}\right]}
\label{eqn-diskdensity}
\end{equation}
where $R_{C}$ is the characteristic radius of the protoplanetary disk and the exponent $\gamma$ sets the slope of the gas surface density distribution. 
The disk is truncated at the user-defined inner edge $R_{in}$ 
and the gas density is set to zero inward of $R_{in}$.

The temperature profile in the disk midplane is characterized by the radial power law of irradiated disks \citep{hayashi1981,isella2016}:
\begin{equation}
T(r) = T_{0} \left(\frac{R}{1\,{\rm au}}\right)^{-\beta}\,K
\label{eqn-disktemperature}
\end{equation}
where $T_{0}$ is the disk temperature at 1 AU and $\beta$ is the slope of the temperature profile, both to be provided as input. 

The density and temperature profiles of the disk from Eqs. \ref{eqn-diskdensity} and \ref{eqn-disktemperature} are used to compute the damping effects of the gas on the dynamical evolution of particles in the n-body simulations following \citet{weidenschilling1977}. The drag acceleration $F_D$ is:
\begin{equation}
F_{D} = \frac{3}{8}\frac{C_{D}}{r_{p}}\frac{\rho_{g}}{\rho_{p}}\Delta v^{2}
\label{eqn-gasdrag}
\end{equation}
where $C_{D}$ is the gas drag coefficient, $\rho_{g}$ is the local volume density of the gas \citep{armitage2010}, $\rho_{p}$ and $r_{p}$ are the average density and physical radius of the planetesimals, respectively 
, and $\Delta v$ is the relative velocity between gas and planetesimals.

The volume density of the gas in the midplane is 
\begin{equation}
\rho_{g}(r)=\frac{1}{\sqrt{2\pi}}\frac{\Sigma(r)}{h}    
\end{equation}
where $h=c_{s}/\Omega$ is the is the vertical disk scale-height, $c_{s}$ and $\Omega$ are the isothermal sound speed and the orbital frequency at the orbital distance $r$ \citep{armitage2010}. The relative velocity $\Delta v$ of the gas with respect to the local Keplerian velocity $v_{k}(r)$ is derived from the computation of the sub-Keplerian gas velocity following \citet{brasser2007}:
\begin{equation}
v_{g}(r) = v_{k}(r)\sqrt{1-2\eta(r)}
\end{equation}
where $\eta(r)$ is
\begin{equation}
\eta(r)=\frac{1}{2\gamma_{C}}\left(\gamma + \beta/2 + 3/2\right)\left(\frac{c_s}{v_k}\right)^{2}
\end{equation}
where $\gamma_{C}=7/5$ is the ratio of the specific heats of molecular hydrogen \citep{brasser2007}.

The gas drag coefficient $C_D$ of each planetesimal is computed following the treatment described by \citet{tanigawa2014} as a function of the Reynolds ($Re$) and Mach ($Ma$) numbers \citep[see][for the definitions adopted]{brasser2007}, coupling the individual drag coefficients to the specific orbit of the planetesimal and to the local disk environment:
\begin{align}
C_{D} = & \left[\left(\frac{24}{Re}+\frac{40}{10+Re}\right)^{-1}+0.23Ma\right]^{-2} \nonumber \\
        & +\frac{2\cdot\left(0.8k+Ma\right)}{1.6+Ma}  
\end{align}
where $k$ is a factor with value of 0.4 for $Re < 10^{5}$ and 0.2 for $Re > 10^{5}$ \citep{tanigawa2014,nagasawa2019}. This formulation of the drag coefficient $C_D$ satisfies the condition of smooth transitions across different drag regimes as discussed by \cite{stoyanovskaya2020}.

After the onset of the runaway gas accretion phase (i.e. for $t > \tau_{c}$), giant planets are  assumed to be surrounded by a gap with width \citet{marzari2018}:
\begin{equation}
W_{gap} = C\cdot R_{H}    
\end{equation}
where the numerical proportionality factor $C=4$ is from \citet{marzari2018}. Following \citet{turrini2021}, the gas density in the gap $\Sigma_{gap}(r)$ evolves over time with respect to the local unperturbed gas density $\Sigma(r)$ as:
\begin{equation}
\Sigma_{gap}(r) = \Sigma(r)\cdot \exp{\left[-\left(t-\tau_{c}\right)/\tau_{g}\right]}
\end{equation}
where $\tau_{c}$ and $\tau_{g}$ are the same as in Eqs. \ref{eqn-coregrowth} and \ref{eqn-gasgrowth}.

The effects of disk gravity on the orbital motion of planetesimals are computed through the analytical treatment for axisymmetric thin disks of \citet{ward1981}, which \citet{fontana2016} showed being in good agreement with the perturbations computed from hydrodynamic simulations. 
Following \citet{nagasawa2019}, the computation focuses on the leading term of the force due to disk gravity (F$_\text{DG}$) which, when $\gamma \ne 1$, is:
\begin{equation}
    F_{DG} = 2 \pi G \Sigma(r) \sum^{\infty}_{n=0} A_{n} \frac{\left(1-\gamma\right)\left(4n+1\right)}{\left(2n+2-\gamma\right)\left(2n-1+\gamma\right)}
\label{eqn-selfgravity}
\end{equation}
where $A_{n}=\left[ (2n)!/2^{2n}(n!)^{2} \right]^{2}$ \citep{ward1981,marzari2018}. When $\gamma=1$, the numerator of the right-hand summation of Eq. \ref{eqn-selfgravity} becomes zero, so the following leading term in the analytical formulation of \citet{ward1981} is computed instead:
\begin{equation}
    F_{DG} = 2 \pi G \Sigma(r) \sum^{\infty}_{n=0} A_{n} \frac{\left(2n+1\right)}{\left(2n+2-\gamma\right)}\left(\frac{R_{in}}{r}\right)^{2n+2-\gamma}
\end{equation}
For all practical purposes $\frac{R_{in}}{r} < 1$ (see below) and the contribution of all terms for $n\geq1$ rapidly goes to zero. Focusing on the leading term $n=0$ the equation simplifies to
\begin{equation}
    F_{DG} = 2 \pi G \Sigma(r)\left(\frac{R_{in}}{r}\right)
\end{equation}

Focusing on the leading terms that determine disk gravity is computationally efficient and has been shown to be accurate as long as the planetary bodies are distant from both the inner and outer edges of the disk \citep{fontana2016}. However, for all practical purposes this approach proves adequate also when this condition is not satisfied \citep{turrini2021}. 

In the inner disk regions, the dynamical evolution of planetesimals is dominated by the damping effects of gas drag that promptly cancel out any excitation due to the disk gravity. Numerical experiments including the additional terms required to account for the effects of the inner disk edge showed no appreciable difference in the dynamical evolution of the planetesimals \citep{turrini2021}. Given the wider extension of the disk gas with respect to solids \citep{mordasini2016,isella2016,ansdell2018,facchini2019}, the only planetary bodies that approach the disk outer edge in typical planet formation simulations are those scattered into an extended disk or ejected from the planetary system by massive planets \citep{polychroni2025}. For these bodies, the dominant contributions that shape their dynamical evolution are the close encounters with the massive planets, while the rapidly decreasing gas density limits the role of disk gravity.

\subsubsection{HPC implementation of the interactions with the gas}\label{sec:hpc-gas}

Modeling the interactions between planetary bodies and disk gas is one of the most computationally demanding task in {\sc Mercury-Ar$\chi$es } and, depending on the simulated planetary architecture, it can outweigh the computation of the Newtonian forces (see Sect. \ref{sect:performance}). The direct computation of Newtonian forces in {\sc Mercury} scales as $O(N_{big} N_{tot})$ where $N_{big}$ is the number of massive bodies, $N_{tot}=(N_{small}+N_{big})$ and $N_{small}$ is the number of massless particles. The interactions with the gas scale linearly on $N_{tot}$ but are characterized by higher arithmetic intensity per body, $\beta$, globally scaling as $O(\beta N_{tot})$. 

When the computational weight of $\beta$ proves greater than that of $N_{big}$, e.g. when simulating few planetary bodies embedded into an extended planetesimal disk as in the use cases in Sect. \ref{sect:performance}, $O(\beta N_{tot})$ grows faster than $O(N_{big}N_{tot})$ and becomes the dominant contribution to the computational load of the simulation. It is therefore critical to ensure that the implementation of the gas-particle interactions benefits from both parallelization and vectorization. {\sc Mercury-Ar$\chi$es} achieves this by pre-computing in separate loops all quantities associated to complex if-branches or non-contiguous stride in memory access that would prevent vectorization. The following code example showcases the identification of massless particles crossing the disk inner edge or the gaps surrounding forming planets, and the computation of multiplicative factors to correct the gas density they experience: 
\begin{verbatim}
!$omp parallel do default(shared) private(i)
      do j = nbig+1,nbod
        if ((Rxyd(j).lt.inner_edge)) then
          dens_factor(j)=0.0d0
        else if (nbig.gt.1.and.growthflag
     &    .eqv..true.) then
          do i = 2,nbig
            if ((growthflag2(i).eqv..true.).and.
     &        (time.gt.tstart2(i))) then
              if ((Rxyd(j).lt.(rpla(i)+gap_w(i)))
     &        .and.(Rxyd(j).gt.(rpla(i)- 
     &        gap_w(i)))) then
                dens_factor(j)=exp(-(time-
     &          tstart2(i))/tfold2(i))
              end if
            end if
          end do
        end if  
      end do
!$omp end parallel do
\end{verbatim}

\subsection{Growth of the forming planets}\label{sec:planetary_growth}

The growth of the planets is modeled using the two-phases parametric approach from \citet{turrini2011,turrini2019}, which allows fitting the growth tracks from physically realistic simulations based on planetesimal accretion \citep{lissauer2009,dangelo2021} and pebble accretion \citep{bitsch2015,johansen2019}.

The first phase reproduces the growth of the planet by accretion of solids and the subsequent capture of an extended primary atmosphere composed of disk gas. The duration of this phase is set by the timescale $\tau_{c}$, during which the planetary mass $M_p$ grows from the initial value $M_{0}$ to the final value $M_{c}$ as
\begin{equation}
 M_{p}=M_{0}+\left( \frac{e}{e-1}\right)\left(M_{c}-M_{0}\right)\left( 1-e^{-t/\tau_{c}} \right)
\label{eqn-coregrowth}
\end{equation}
The physical radius of the growing planet $R_{p}$ evolves alongside the planetary mass following the approach described by \citet{lissauer2009} and \citet{fortier2013}:
\begin{equation}
 R_{p} = \frac{G\,M_p}{c_{s}^{2}/k_{1}+\left(G\,M_{p}\right)/\left(k_{2}R_{H}\right)}
 \label{eqn-inflatedradius}
\end{equation}
where $G$ is the gravitational constant, $M_p$ and $R_H$ are the instantaneous mass and Hill's radius of the planet, respectively, $c_{s}$ is the sound speed in the protoplanetary disk at the orbital distance of the planet, and $k_{1}=1$ and $k_{2}=1/4$ are constants \citep{lissauer2009}.
The parameters $\tau_c$, $M_0$ and $M_c$ are input values of the simulation.

The second phase reproduces the runaway gas accretion that allows the forming planet to become a gas giant. The mass growth during the runaway gas accretion is modeled as:
\begin{equation}
 M_{p}=M_{c}+\left( M_{f} - M_{c}\right)\left( 1-e^{-(t-\tau_{c})/\tau_{g}}\right)
\label{eqn-gasgrowth}
\end{equation}
where $M_{f}$ is the final planetary mass while $\tau_{g}$ is the e-folding time of the runaway gas accretion. During the gravitational infall of the gas, the planetary radius shrink over time as:
\begin{equation}
 R_{p} = R_{c} - \Delta R \left(1-\exp^{-(t-\tau_{c})/\tau_{g}}\right)
 \label{eqn-collapsingradius}
\end{equation}
where $R_{c}$ is the planetary radius at $\tau_{c}$, i.e. the onset of the runaway accretion, and $\Delta R = R_{c} - R_{I}$ is the decrease of the planetary radius during the gravitational collapse of the gas. The parameters $\tau_g$, $M_f$ and $\Delta R$ are input values of the simulation. 

\subsection{Migration of the forming planets}\label{sec:orbital_migration}

The migration of the forming planets due to their interactions with the surrounding protoplanetary disk is modeled through the two-phases approach from \citet{turrini2021}, which in turn semi-analytically models the realistic non-isothermal migration tracks from the population synthesis models by \citet{mordasini2015}. The two migration phases are characterized by linear and power-law migration regimes, respectively, and are implemented based on the analytical treatment of \citet{hahn2005}. 

The first phase is described by a linear migration regime governed by the drift rate $\Delta v_{1}$ \citep{turrini2021}:
\begin{equation}
    \Delta v_{1} = \frac{1}{2}\frac{\Delta a_{1}}{a_{p}}\frac{\Delta t}{\tau_{1}}v_{k}
\label{eqn-typeImigration}
\end{equation}
where $\Delta a_{1}$ and $\tau_{1}$ are the radial displacement and duration of this migration phase, $v_{k}$ and $a_{p}$ are the Keplerian velocity and semimajor axis of the planet, and $\Delta t$ is the timestep used in the n--body simulations. 

The second migration phase is described by a power law migration regime governed by the drift rate $\Delta v_{2}$ \citep{hahn2005}:
\begin{equation}
    \Delta v_{2} = \frac{1}{2}\frac{\Delta a_{2}}{a_{p}}\frac{\Delta t}{\tau_{2}}\exp^{-\left(t-\tau_{1}\right)/\tau_{2}} v_{k}
\label{eqn-typeIImigration}
\end{equation}
where $\Delta a_{2}$ and $\tau_{2}$ are the radial displacement and duration of this migration phase, while all other parameters are the same as those in Eq. \ref{eqn-typeImigration}.

From a physical point of view, the two migration regimes are coupled to the two growth phases discussed in Sect. \ref{sec:planetary_growth} meaning that $\tau_{1}=\tau_{c}$ and $\tau_{2}=\tau_{g}$ \citep[][]{turrini2021,turrini2023}. The implementation of the two migration regimes, however, allows for decoupling orbital migration from planetary growth and introduce arbitrary user-defined migration histories, e.g. by delaying the onset of migration with respect to the onset of solid growth or extending the linear migration regime also during the gas accretion phase.

\subsection{Planetesimal disk}\label{sect-planetesimal_disk}

{\sc Mercury-Ar$\chi$es} models planetesimal disks as swarms of dynamical tracers possessing inertial mass but no gravitational mass. The lack of gravitational mass means that tracers do not affect each other nor the planets, behaving as massless particles at population level. The attribution of inertial mass allows for quantifying the way planetesimals are affected by the disk gas through the processes discussed in Sect. \ref{sect-protoplanetary_disk}.

The inertial mass is quantified from the physical radius and the density of the planetesimals. The current implementation of {\sc Mercury-Ar$\chi$es} allows for attributing in input the global characteristic radius of all planetesimals 
The selection of the planetesimal density is currently hard-coded in {\sc Mercury-Ar$\chi$es}: 
tracers whose initial semimajor axes position them inward of the water snowline are considered rocky and attributed an average density of $\rho_{rock}$=2.4 g cm$^{-3}$ (\citealt{turrini2014c} based on asteroid densities from \citealt{britt2002,carry2012}). Tracers whose initial semimajor axes position them outward of the water snowline are considered ice-rich and attributed an average density $\rho_{ice} = 1$ g cm$^{-3}$ following \citet{turrini2019}. This value mediates between cometary densities (0.4-0.6 g cm$^{-3}$, see \citealt{brasser2007} and references therein and \citealt{jorda2016}) and the density of the 200 km-wide ice-rich Saturnian irregular satellite Phoebe (1.63 g cm$^{-3}$, \citealt{porco2005}).

\subsection{Updated ejection criterion for planetary bodies}

{\sc Mercury-Ar$\chi$es} implements the physically-justified ejection criterion where planetary bodies are flagged for ejection when their orbits become physically unbound (i.e., the orbital eccentricity is greater or equal than unity; \citealt{polychroni2025}), in place of the original criterion of {\sc Mercury} based on threshold orbital distances specified in input by the users. As the computation of the orbital elements is more expensive than that of the instantaneous radial distance, we parallelized the block of instructions as shown below:

\begin{verbatim}
!$omp  parallel do simd default(shared)
|$omp& private(gm,q,ecc,inc,tmp1,tmp2,tmp3)
      do j = i0, nbod
        if (j.le.nbig) then
          gm=m(1)+m(j)
        else
          gm=m(1)
        end if
        call mco_x2el (gm,x(1,j),x(2,j),x(3,j),
     &  v(1,j),v(2,j),v(3,j),q,ecc,inc,tmp1,
     &  tmp2,tmp3)
        if (ecc.ge.1.0d0) then
          stat(j) = -3
          m(j) = 0.d0
          s(1,j) = 0.d0
          s(2,j) = 0.d0
          s(3,j) = 0.d0
        end if
      end do
!$omp end parallel do simd    
\end{verbatim}

\subsection{Migration and time-step adapting}

Depending on the adopted migration track, the forming planets in the simulations may reach the orbital regions close to the host star where correctly resolving the temporal evolution of the orbits requires extremely small timesteps. Adopting the necessary small timestep across the whole duration of the simulation would prove prohibitively expensive from a computational point of view. To effectively address this issue, {\sc Mercury-Ar$\chi$es} allows the user to start the simulation with the optimal timestep for the input architecture and switch to the smaller timestep when required by automatizing the approach adopted by \citet{juric2008}.

When the planet approaches the orbital distance where the initial timestep proves too long to correctly sample the orbit, the n-body code decreases the timestep to 1/25 of the orbital period at the minimum semimajor axis specified in input. While the high numerical stability of the {\sc WHFAST} Kepler solver \citep{rein2015} is expected to ensure the smooth transition between the two timesteps, for increased accuracy {\sc Mercury-Ar$\chi$es} applies the hybrid approach of {\sc Mercury} also during the timestep when the transition in $dt$ occurs and computes the Keplerian orbital motion of the bodies requiring the updated timestep using the Bulirsh-Stoer integrator.

\subsection{Temporal evolution of the planetary system}

While the original output files of {\sc Mercury} are preserved for compatibility with existing legacy codes, {\sc Mercury-Ar$\chi$es} introduces new output files in the form of sequential snapshots of the dynamical and physical state of the planetary system. During runtime, these files serve as quick-look windows on the real-time state of the simulated system. At the end of the simulations, they support the analysis of its global temporal evolution without the need of unpacking {\sc Mercury}'s output, which stores the evolution of each planetary body in individual files and requires processing an impractically large number of files in modern planet formation simulations.

The snapshots are timed by the same frequency of the data dumps for the creation of restart files, to guarantee the alignment of the outputs in the case of interruptions or when restarting the simulations.
The snapshot files report the time of the snapshot in simulation time followed by the orbital information of all bodies, massive and massless. For all bodies the snapshots report the osculating main orbital elements - semimajor axis, eccentricity and inclination - followed by the initial semimajor axis $a_0$, used to trace the formation region of the bodies, and the state vectors 
$\Vec{x}$ and $\Vec{v}$. For the massive planets, the snapshots also report their radius and mass values.

\section{Workload profile and code performance}\label{sect:performance}

{\sc Mercury-Ar$\chi$es} is designed as a flexible investigation tool capable of adapting to the large diversity of observed architectures of planetary systems and of their stellar and disk formation environments \citep[e.g.][]{Zhu2021,Feinstein2025}. This adaptability, on the other hand, means that workload profiles are system-dependent and there is no unique profile on which to tune {\sc Mercury-Ar$\chi$es}'s performance. When modeling planetary systems without migrating planets, in absence of large-scale dynamical instabilities close encounters will be comparatively sparse over the simulation duration and most of the workload will reside in the parallel regions. When modeling systems where planets undergo migration, close encounters will occur more frequently and will increase the relative computational load of the serial parts of the code, without necessarily activating their conditional parallelization. Moreover, as introduced in Sect. \ref{sec:hpc-gas}, depending on the relative numbers of massive and massless bodies in each simulation the workload profile can be dominated by the direct computation of the Newtonian forces or by the gas-bodies interactions.

In the following, we will focus on use cases from recent exoplanetary investigations performed with {\sc Mercury-Ar$\chi$es} to illustrate and discuss the performance of the code. Specifically, we will adopt two use cases where giant planets form and migrate across extended planetesimal disks reaching compact final orbits around their host stars. In both use cases, the giant planets start as lunar-mass planetary embryos (0.01 M$_\oplus$) to grow to Jupiter-like masses (100-300 M$_\oplus$), while the planetesimal disks are represented by swarms of massless particles embedded in protoplanetary disks with mass equal to 5\% that of their central stars and interacting with the surrounding gas. The first use case will be used to analyze the performance of {\sc Mercury-Ar$\chi$es} on consumer-grade infrastructures, while the second will be used to analyze its performance and scalability on high-end HPC nodes. Further details on the profiling of the code are provided in Simonetti et al., this issue, while information on the code runtime in the campaign of simulations of the OPAL project is provided in Polychroni et al., this issue.

\subsection{Use case 1: Intel Performance Hybrid Architectures}\label{sec:use_case1}

\begin{figure}[t]
    \centering
    \includegraphics[width=\columnwidth]{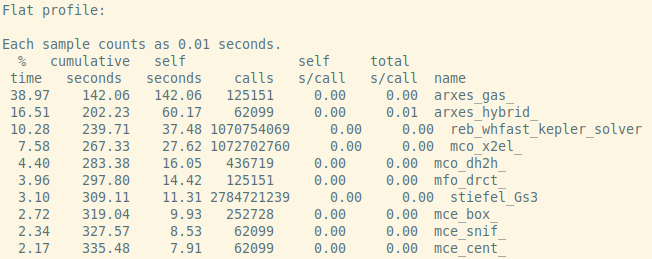}
    \caption{Flat profile produced by {\sc Gprof} for the serial run of {\sc Mercury-Ar$\chi$es} described in Sect. \ref{sec:use_case1}. The simulated planetary system hosted two growing and migrating planets and about 18000 massless particle interacting with the disk gas. The flat profile shows the ten most computationally expensive subroutines.}
    \label{fig:gprof-profile}
\end{figure}

\begin{figure}[b]
    \centering
    \includegraphics[width=\columnwidth]{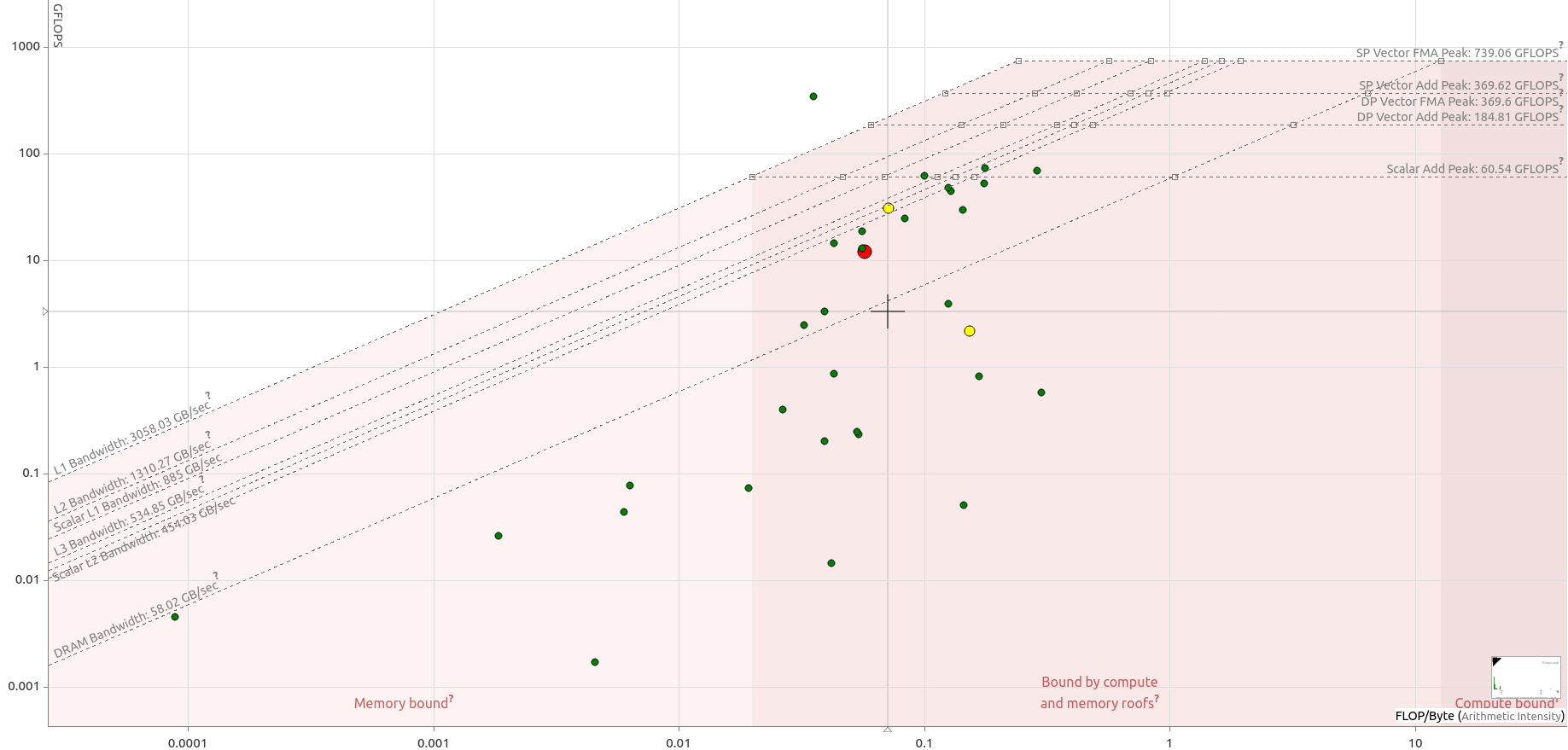}
    \caption{Roofline analysis produced by Intel Advisor for the parallel run of {\sc Mercury-Ar$\chi$es} described in Sect. \ref{sec:use_case1} using the same planetary system as Fig. \ref{fig:gprof-profile}. The red symbol is the most computationally intensive subroutine, {\sc arxes\_gas}, while the two yellow ones are {\sc mco\_x2el} and {\sc mce\_cent} (see discussion in Sect. \ref{sec:use_case1} for details). The roofline plot highlights how most of the computational load of the simulation resides in the compute and memory bound region.}
    \label{fig:roofline-profile}
\end{figure}

The first analysis is performed on an HP Z2 G9 workstation running Linux Mint 21.3 and equipped with 12th Generation Intel Core i9-12900 and 64 GB of DDR5-4800 RAM. This processor is based on Intel Performance Hybrid Architecture and integrates two types of cores into a single die\footnote{\url{https://www.intel.com/content/www/us/en/products/sku/134597/intel-core-i912900-processor-30m-cache-up-to-5-10-ghz/specifications.html}}: Performance-cores (P-cores) and Efficient-cores (E-cores). P-cores are designed to maximize processing power, measured as instructions per cycle, and are capable of hyper-threading. E-cores are designed instead to maximize energy efficiency, measured as performance-per-watt, and are not capable of hyper-threading. The i9-12900 processor is equipped with 8 P-cores (physical cores: id. 0-7, hyper-threading virtual cores: id. 8-15) capable of turbo frequencies up to 5 GHz and 8 E-cores (physical cores: id. 16-23) capable of turbo frequencies up to 3.8 GHz. Performance hybrid architecture processors are equipped with two additional technologies that impact scalar and multi-thread performance: Intel Thread Director and Intel Turbo Boost Max 3.0. Intel Thread Director is capable of automatically routing the workload to the most efficient cores, hence dynamically optimizing the global workflow. Intel Turbo Boost Max 3.0, alongside boosting the clock frequencies of all cores, is capable of identifying up to two of the fastest cores on the processor and further boost their clock frequencies. The combination of these technologies with the Performance Hybrid Architecture of the processor may results in deviations between the performance of the code when run serially and with few threads with respect to when it is run with higher thread counts.

Based on the suggested flags of the PRACE Best Practice Guides\footnote{\url{https://prace-ri.eu/resources/documentation/best-practice-guides/}} \citep{PRACE2020}, {\sc Mercury-Ar$\chi$es} is compiled with {\sc Gfortran} 11 using the flags {\sc -std=legacy -Ofast -march=core-avx2 -mtune=core-avx2 -malign-data=cacheline -flto} (plus {\sc -fopenmp} when run in mulit-threaded mode), while {\sc WHFAST} is compiled with {\sc GCC} 11 using the flags {\sc std=c99 -O3 -march=core-avx2 -mtune=core-avx2 -flto}. The use case adopted is based on one of the scenarios (id. 3) simulated by \cite{turrini2023} and considers two forming giant planets starting respectively at 10 and 20 AU from the host star and a planetesimal disk extending from 1 to 25 AU composed of 1000 massless particles/au, for a total of about 18000 particles. We refer interested readers to \citet{turrini2023} for additional details on the simulated planetary architecture and formation history. Since the use case focuses on the code performance rather than scientific accuracy, we considered simulation durations as well as formation and migration timescales 100-200 times shorter than in the original simulations from \citet{turrini2023}, meaning that each simulation spans $10^4$ years, the planetary cores grow in $5\times10^3$ years and they undergo runaway gas accretion with a characteristic timescale of $10^3$ years. This setup results in a serial runtime of 3703 seconds. 

Before proceeding with the simulations, we performed two preliminary test runs, one serial and one running on eight thread on OpenMP, with timescales further reduced by a factor of 10, for a serial runtime of 364.54 seconds. The serial run was profiled with {\sc Gprof} and produced the flat profile of Fig. \ref{fig:gprof-profile}, which shows the ten most computationally expensive subroutines. As discussed in Sect. \ref{sec:hpc-gas}, this use case is dominated by the computational cost of the interactions of the planetary bodies with the gas ({\sc arxes\_gas}). The parallel run, where we let Intel Thread Director handle the distribution of the workload, was used to produce the roofline analysis of Fig. \ref{fig:roofline-profile} using Intel Advisor. The resulting roofline plot reveals that the majority of the computational load of the simulation falls in the region characterized by both memory and compute bounds. The most computationally heavy subroutine remains {\sc arxes\_gas} (red symbol in Fig. \ref{fig:roofline-profile}), followed by the conversion of state vectors into orbital elements ({\sc mco\_x2el}, top yellow symbol and the fourth entry in the flat profile of Fig. \ref{fig:gprof-profile}) and the computation of encounters with the central star ({\sc mce\_cent}, bottom yellow symbol and the tenth entry in the flat profile of Fig. \ref{fig:gprof-profile}). The latter two subroutines are implemented serially, but the calls to {\sc mco\_x2el} have been inserted wherever possible into parallel OpenMP {\sc DO} loops, thus optimizing their use in the simulations.

\begin{figure}
    \centering
    \includegraphics[width=\columnwidth]{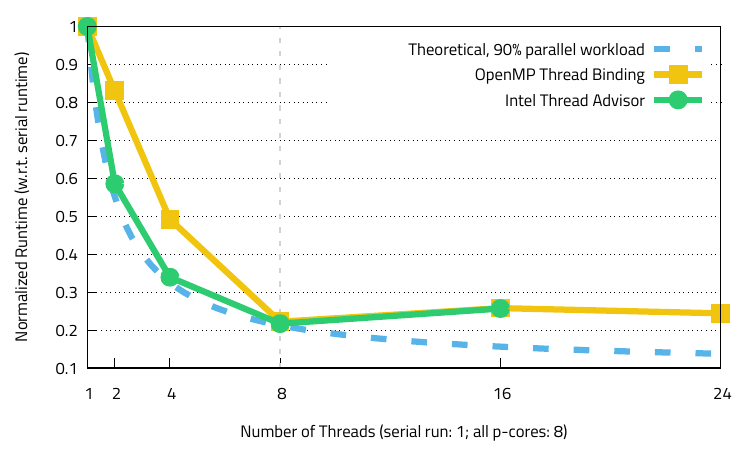}
    \caption{Comparison of the runtime values of {\sc Mercury-Ar$\chi$es} in the tests of Sect. \ref{sec:use_case1} when running serially, in parallel with no thread binding (i.e. leaving the management of the workload schedule to the processor and the operating system) and in parallel with thread binding. All runtime values are normalized to the duration of the serial run. The theoretical dashed curve shows the expected runtime based on Amdahl's law when 90\% of the workload is parallelized. The vertical dashed line marks the number of P-cores available on the processor.}
    \label{fig:i9-runtime}
\end{figure}

We performed six simulations with the nominal setup adopted for this use case: a serial one and five parallel ones using 2, 4, 8, 16 and 24 cores. The first three parallel simulations run on P-cores, with Intel Thread Director deciding which cores to use between the physical and virtual ones. The simulation with 16 and 24 cores simultaneously make use of both P-cores and E-cores, with the case running on 24 cores also using hyper-threading. To test the computational implications of Performance Hybrid Architectures, we performed two sets of parallel simulations: the first set did not use any thread binding and relied on the operating system and Intel Thread Director to manage the workload distribution, while the second set implemented thread binding using the environmental variables {\sc OMP\_PROC\_BIND="TRUE"} and {\sc OMP\_PLACES="THREADS"}. In the simulations with 2, 4 and 8 threads we forced the use of the P-cores setting {\sc GOMP\_CPU\_AFFINITY=0-7}, while in the simulation with 16 threads we forced the use of P-cores and E-cores setting {\sc GOMP\_CPU\_AFFINITY=0-7,16-23}. The results of the two sets of simulations are compared in Fig. \ref{fig:i9-runtime} in terms of their normalized runtime with respect to the serial run. Fig. \ref{fig:i9-runtime} also show the theoretical normalized runtime curve from Amdahl's law when 90\% of the total computational workload is parallelized.

The performance of the simulations running without thread binding follows exactly the theoretical curve with 90\% parallel workload until all P-cores are used, while the simulations performed with thread binding using 2 and 4 cores visibly deviate from the theoretical expectations. Observing the core usage profile during runtime with and without thread binding hints to the explanation for these deviations. When the processor and the operating systems are allowed to decide which cores to use, the simulation with two threads run exclusively on two specific cores (virtual cores id. 8 and 13), likely those identified as top performing by Intel Turbo Boost Max 3.0. When thread binding is implemented, the specified pool of cores (physical cores id. 0-7) does not allow Intel Turbo Boost Max 3.0 to provide the additional frequency boost, resulting in lower performance with respect to the serial and unbounded parallel run. In the case of the simulation with 4 threads, the processor and operating system shift the workload between cores in a round-robin fashion but less frequently than observed in pre-Performance Hybrid Architecture processors. Furthermore, the top performing cores (id. 8 and 13) are used as often as possible but not in a continuous way. A plausible explanation is that the new technologies present on the processor successfully optimize the use of boosted frequencies and the thermal balance of the cores, while in the simulation run with thread-binding this is not feasible, resulting in degraded performance. The simulations running with 16 and 24 threads show lower performance than the case with 8 threads, with Intel Thread Director causing the simulation with 24 threads to stall without completing. This performance loss is likely caused by the compute and memory bound nature of the code, specifically the lower computational gain provided by the E-cores not  compensating for the allocation of cache memory to the individual threads.

\subsection{Use case 2: Leonardo Pre-Exascale Infrastructure}\label{sec:use_case2}

\begin{table}[t]
    \centering
    \begin{tabular}{l c c c }
    \hline
     Particles  & Execution & Runtime (hours) & Speedup\\
    \hline
    \multirow{2}{*}{10000}  &   Serial     & 86.07 & \multirow{2}{*}{4.76}\\
                            &   Parallel   & 18.07\\
    \hline
    \multirow{2}{*}{50000}  &   Serial     & 290.6 & \multirow{2}{*}{5.94}\\
                            &   Parallel   & 48.86 & \\
                            
    \hline
    \end{tabular}
    \caption{Comparison of the serial and parallel runtime of {\sc Mercury-Ar$\chi$es} on one node of Leonardo's DCGP module for the test cases in Sect. \ref{sec:use_case2} when simulating 1000 particles/au and 5000 particles/au. The parallel simulations are run with 14 threads using the version of {\sc Mercury-Ar$\chi$es} containerized with Singularity adopted by the OPAL project (Polychroni et al., this issue). {\sc Mercury-Ar$\chi$es} is compiled with Intel OneAPI using the compilation flags described in Sect. \ref{sec:use_case2}. The speedups of the parallel runs are consistent with 85-90\% of the computational load being run in parallel. The average runtime/particle provides a measure of the computational efficiency of the code and shows that it improves moving from 1000 to 2000 particles/au and remains mostly constant for increasing numbers of particles.}
    \label{tab-runtimes}
\end{table}

The second use case is run on the Leonardo Pre-Exascale Infrastructure\footnote{\url{https://leonardo-supercomputer.cineca.eu/hpc-system/}} using one node of its Data Centric General Purpose (DCGP) module. Each DCGP node is equipped with two 56-core Intel Xeon Platinum 8480+ CPUs and 
512 GB of DDR5-4800 RAM, with each core capable of turbo frequencies up to 3.8 GHz\footnote{\url{https://www.intel.com/content/www/us/en/products/sku/231746/intel-xeon-platinum-8480-processor-105m-cache-2-00-ghz/specifications.html}} . The use case follows the formation and migration of the giant planet WASP-69b around its host star and is derived from  the \textit{Origins of Planets for ArieL} (OPAL) project on Leonardo (Polychroni et al., this issue) in support of the ESA space mission Ariel \citep{tinetti2018}. The giant planet starts the simulation as a lunar-mass planetary embryo (0.01 M$_\oplus$) located at 10 AU and ends its formation process at 0.4 AU with a mass of 81.05 M$_\oplus$. During its growth and migration, the forming planet interacts with a planetesimal disk extending from 1 AU to 12.5 AU. While interacting with the forming planet, the planetesimal disk is affected by the protoplanetary disk within which it is embedded both in terms of aerodynamic drag and disk gravity. The host star has mass of 0.81 M$_\odot$ while the mass of the protoplanetary disk is 5\% of the stellar mass. 
The timestep of the simulations is 18 days and each simulation spanned a total duration of 1.5 Myr.

The setup of this use case implies that the forming planet interacts with a large number of planetesimals at each timestep, meaning that the {\sc critical} code region responsible for the identification of the close encounters is called at most timesteps. Furthermore, at different timesteps the close encounters may be more effectively handled in serial than in parallel, meaning that the parallel fraction of the computational load fluctuates during runtime. Finally, the inclusion of only one forming planet reduces the computational weight of the Newtonian forces with respect to the first use case. To test how these characteristics of the use case affect performance and scalability, we run a set of parallel simulations with different number of particles and two serial ones. All simulations were run on a single node of the DCGP module of Leonardo, with {\sc Mercury-Ar$\chi$es} being compiled with Intel OneAPI using the flags {\sc -O3 -march=core-avx2 -align array64byte -fma -ipo} (plus {\sc -qopenmp} when run in multi-threaded mode), while {\sc WHFAST} is compiled using the flags {\sc -std=c99 -O3 -march=core-avx2 -fma -ipo}. We considered four simulations using 1000, 2000, 5000 and 10000 particles/au to represent the planetesimal disks, resulting in total numbers of planetesimals of about 10000, 20000, 50000, 100000 massless particles. Each set of four simulations was run serially and in parallel with thread-binding using 14 threads per simulation. The parallel simulations were run with the same version of {\sc Mercury-Ar$\chi$es} containerized with Singularity used in OPAL.

\begin{figure}[t]
    \centering
    \includegraphics[width=\hsize]{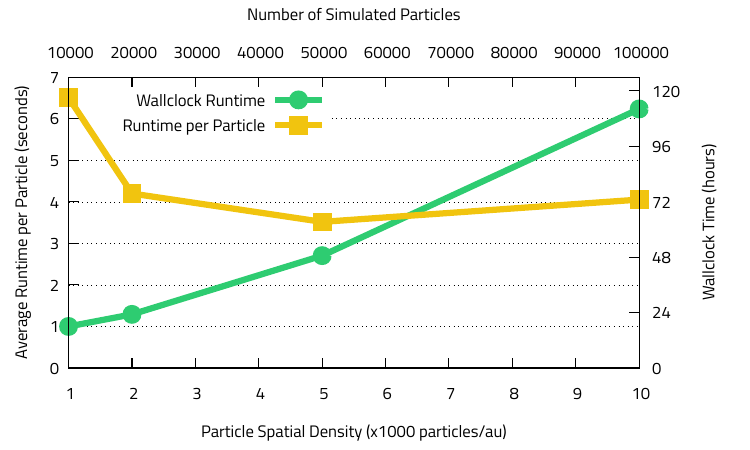}
    \caption{Comparison of the runtime of the parallel simulations of {\sc Mercury-Ar$\chi$es} in Sect. \ref{sec:use_case2}. The simulations are run on one node of Leonardo's DCGP module using 14 threads and considering planetesimal disks extending by about 10 AU and populated by 1000, 2000, 5000 and 10000 particles/au. The green curve with circle symbols shows the evolution of the wallclock runtime, the golden curve with square symbols shows the average runtime per particle. The parallel simulations are run with the version of {\sc Mercury-Ar$\chi$es} containerized with Singularity adopted by the OPAL project (Polychroni et al., this issue). {\sc Mercury-Ar$\chi$es} is compiled with Intel OneAPI using the compilation flags described in Sect. \ref{sec:use_case2}.
    }
    \label{fig-performance}
\end{figure}

The wallclock runtime the serial and parallel simulations considering 10000 and 50000 particles are reported in Table \ref{tab-runtimes}. Their comparison reveals speedups of about a factor of 5 between serial and parallel simulations, with the run with 50000 particles proving overall more efficient. Using Amdahl's law, the speedups reported in Table \ref{tab-runtimes} are consistent with 85\% of the workload being parallelized in the simulation with 10000 particles, while this percentage grows to about 90\% in the simulation with 50000 particles. Fig. \ref{fig-performance} 
shows the comparison of the total runtime (green curve with circles) and the average runtime per particle (golden curve with squares) of the four parallel simulations as a function of their increasing workload. The simulation with 10000 particles is the least efficient due to insufficient workload, in agreement with the information provided by Table \ref{tab-runtimes}. The average runtime per particle decreases when the computational workload increases: doubling the number of particles results in a limited runtime increase of about 30\%, meaning that the overall computational efficiency grows. The normalized runtime/particles values in Fig. \ref{fig-performance} 
confirm that the computational efficiency remains almost constant for larger workloads, the case with 50000 particles providing the best performance as discussed for Table \ref{tab-runtimes}.


\section{Conclusions}
 
In this work we described the high-performance implementation of the planet formation code {\sc Mercury-Ar$\chi$es} and its underlying parallel n-body engine {\sc Mercury+}, which leverages on the OpenMP directive-based parallelism for shared memory environments to take advantage of the multi-thread and vectorization capabilities of modern processors. As we shown in summarizing its modeling capabilities and validating its performance in real use cases, {\sc Mercury-Ar$\chi$es} provides a computationally efficient and physically realistic planet formation platform to simulate the multiple interactions between forming planets and their surrounding protoplanetary disk environments. 

{\sc Mercury-Ar$\chi$es} is available to the community through collaborations and is part of the {Ar$\chi$es} suite of star and planet formation codes being used by the Key Science Project \textit{Origins of Planets for ArieL} (OPAL) on the Leonardo Pre-Exascale Infrastructure (Polychroni et al., this issue) in support of ESA's Ariel space mission \citep{tinetti2018}. In the framework of the KSP OPAL, {\sc Mercury-Ar$\chi$es} is also the foundation for {\sc Mercury-OPAL}, its porting to GPU computing based on the OpenACC directive-based paradigm for heterogeneous computational infrastructures (Simonetti et al., this issue). 

{\sc Mercury+}, the parallel n-body engine of {\sc Mercury-Ar$\chi$es}, is available to the scientific community as stand-alone code upon request to the developing team, providing an enhanced, high-performance version of the widely used {\sc Mercury} code \citep{chambers1999}. The design of both codes ensures seamless compatibility with existing additional n-body libraries, e.g. the DPI library to model planetary systems in binary star systems \citep{DPI2015,Nigioni2025}, and with any legacy code developed by the community to interact with {\sc Mercury}.

\section*{Acknowledgments}

The authors thank the original author of {\sc Mercury} John E. Chambers for providing the foundation for {\sc Mercury+}, and Francesco Marzari, Patricia Verrier, Nader Haghighipour, Sin-iti Sirono, Martina Vicinanza and Eduard Vorobyov, as well as many participants to the MIAPbP workshop ``Planet Formation: From Dust Coagulation to Final Orbit Assembly'' held in 2022 in Garching, for the feedback and interactions that allowed the enhancement and validation of {\sc Mercury-Ar$\chi$es} over the years. This work is supported by the Fondazione ICSC, Spoke 3 “Astrophysics and Cosmos Observations'', National Recovery and Resilience Plan (Piano Nazionale di Ripresa e Resilienza, PNRR) Project ID CN\_00000013 “Italian Research Center on High-Performance Computing, Big Data and Quantum Computing'' funded by MUR Missione 4 Componente 2 Investimento 1.4: Potenziamento strutture di ricerca e creazione di “campioni nazionali di R\&S (M4C2-19)'' - Next Generation EU (NGEU). 
The authors acknowledge support from ASI-INAF grant no. 2021-5-HH.0 plus addenda no. 2021-5-HH.1-2022 and 2021-5-HH.2-2024 and grant no. 2016-23-H.0 plus addendum no. 2016-23-H.2-2021, from the INAF-IFSI Basic Research projects \textit{High Performance Planetology} (HPP), \textit{High Performance Planetology - Second Edition} (HPP-2E), the INAF PRIN GENESIS-SKA and PLATEA, the INAF Main Stream project “Ariel and the astrochemical link between circumstellar discs and planets” (CUP: C54I19000700005), and the European Research Council via the Horizon 2020 Framework Programme ERC Synergy “ECOGAL” Project GA-855130. The authors also acknowledge the support of Amazon Web Services in the form of time allocation on their AWS EC2 infrastructure in the framework of the INAF-ICT2018 project. The authors wish to thank Vega Forneris, Francesco Reale and Mirko Riazzoli for their support in managing the HPP and Genesis computational clusters at INAF, the development and validation platforms of {\sc Mercury-Ar$\chi$es}. This research has made use of the Astrophysics Data System, funded by NASA under Cooperative Agreement 80NSSC21M00561




\bibliographystyle{elsarticle-harv} 
\bibliography{Bibliography.bib}






\end{document}